# Coronavirus RNA Sensor Using Single-Stranded DNA Bonded to Sub-Percolated Gold Films on Monolayer Graphene Field-Effect Transistors

Nicholas E. Fuhr, Mohamed Azize, David J. Bishop

Division of Materials Science and Engineering – 8 St. Mary's St., Boston, Massachusetts 02215

Corresponding author: Nicholas E. Fuhr – fuhrnick@bu.edu

## Abstract:

Electrical detection of messenger ribonucleic acid (mRNA) is a promising approach to enhancing transcriptomics and disease diagnostics because of its sensitivity, rapidity, and modularity. Reported here is a fast SARS-CoV-2 mRNA biosensor (<1 minute) with a limit of detection of 1 aM, and dynamic range of 4 orders of magnitude and a linear sensitivity of 22 mV per molar decade. These figures of merit were obtained on photoresistlessly patterned monolayer graphene field-effect transistors (FETs) derived from commercial four-inch graphene on 90 nm of silicon dioxide on p-type silicon. Then, to facilitate mRNA hybridization, graphene sensing mesa were coated with an ultrathin sub-percolation threshold gold film for bonding 3'-thiolated single-stranded deoxyribonucleic acid (ssDNA) probes complementary to SARS-CoV-2 nucleocapsid phosphoprotein (N) gene. Sub-percolated gold was used to minimize the distance between the graphene material and surface hybridization events. The liquid-transfer characteristics of the graphene FETs repeatedly shows correlation between the Dirac voltage and the copy number of polynucleotide. Ultrathin percolated gold films on graphene FETs facilitate two-dimensional electron gas (2DEG) mRNA biosensors for transcriptomic profiling.

The central dogma postulates that data stored in deoxyribonucleic acid (DNA) structures are accessed by transcription into a complementary ribonucleic acid (RNA) sequences.[1] RNA transcripts may be messengers (mRNAs) or non-coding (ncRNAs) and are able to initiate and regulate gene/protein expression/function.[2] The transcriptome is defined by the summation of all RNA transcripts present at a given moment.[3] Transcriptomics can be read at the cellular, tissue, and population levels to yield information instrumental for the diagnosis and treatment of disease, and ecological insight through elucidating the deep connections underlying biological regulation.[4-8]

Contemporary transcriptomics rely on microarrays and high-throughput RNA sequencing (RNA-Seq).[9] These techniques are reliable but are ultimately limited by costly optoelectronics of complex instruments used to measure fluorescent measurands.[10,11] As a result of the shot noise, signal-to-noise ratios for RNA transcripts at incredulous dilutions (*e.g.*, parts per trillion) are too low to resolve with certainty and therefore remain undetected when using contemporary transcriptomics methods unless amplified in a polymerase chain reaction prior.[12]

It is worthwhile to capture the data regarding these incredulously diluted RNA transcripts to completely resolve transcriptomes, especially as transcriptomics of single-cells, tissues, and individuals becomes more routine. Additionally, rapid resolution of incredulously diluted RNA transcripts would have implications in curbing emergent infectious diseases (EIDs) by augmenting existing PCR technologies used for surveying disease and prompting case management.[13] Given the recent literature landscape regarding novel biosensors, the prospect of combining microarrays and RNA-Seq technologies with two-dimensional (2D) sensor arrays is high and attractive.

The work outlined here aims to prescribe an easy, repeatable method of sensing incredulously dilute RNA transcripts with 2D materials to complement contemporary transcriptomics technologies. 2D, zero-bandgap materials, like graphene, are nearly entirely surface and have properties susceptible to perturbation from the surface environment. For example, the Fermi energy can be elucidated with transfer characteristics, where a gate voltage of minimum conduction defines graphene's charge neutrality point (*i.e.*, the Dirac voltage) and gives insight as to the majority charge carrier type at electrical equilibrium.[14] Using clean, highly crystalline graphene will bring the Dirac voltage towards 0 volts and provides high charge carrier mobility.

Usually, graphene field-effect transistors (FETs) are patterned with photolithography techniques and collect residue throughout the process (*e.g.,* from poly(methyl methacrylate)) and require further treatment to purify the device only partially.[15] Residue can alter the graphene FET performance by increasing charge carrier scattering, lowering charge carrier mobility, and doping the charge carrier concentrations. At the academic level, photoresistlessly patterned graphene FETs are an attractive approach to retaining the graphene's intrinsic properties, without dirtying the surface during the fabrication process, to produce reliable, quasi-pure devices.[16] Here, photoresistless metallization, and subtractive manufacturing of four-inch commercial monolayer graphene/$SiO_2$/p-type Si was facilitated with stainless steel shadow masking techniques and dry, directional plasma etching for yielding FETs for sensing applications.

Over the last decade, the scientific community has realized many graphene-based biosensors with purported sensitivity and selectivity for a plethora of biomolecules. For example, monolayer graphene FETs have been used sensing for heavy metal cations, small molecules like acetylcholine or antibiotics, and larger proteins like virus antigen.[17-22] These examples rely on aptamers or proteins to facilitate molecular recognition and capture but depend on structural confirmation to retain binding function. Solutal variables influence biomolecular structure and function.[23,24]

Other graphene biosensor articles, however, rely on polynucleotide hybridization for electrical detection.[25,26] Alafeef *et al.* used graphene nanoplatelets on paper substrate for their transducer, and relied on well-known gold-thiol chemistry for efficaciously functionalizing their sensor with oligonucleotide-conjugated gold nanoparticles specific to SARS-CoV-2 N gene.[27] Gold is an attractive filming material because it is reactive with free sulfhydryl, a method to reliably anchor biomolecules for biosensing applications.[28] Gold thin films on graphene for reaction with free thiol readily facilitates bonding to molecular recognition elements.

Sub-percolated gold manifests at ultrathin thicknesses and has been reported on graphene/$SiO_2$.[29,30] These gold films are electronically connected to the graphene, doping it, while offering physical protection from ambient molecules (*e.g.,* vaporized water) and other promiscuous signals. The discontinuities expose the material to the environment for sensing applications. Depositing sub-percolated gold films on graphene FETs are shown here to increase the sheet resistance up to a critical thickness ($t_c$) and can be observed in **Figure 1a**.

To leverage this phenomena, 3'-thiolated ssDNA probes complementary to a sequence in the SARS-CoV-2 nucleocapsid phosphoprotein (N) gene were prescribed by the literature and obtained commercially.[27] The probes were reduced and then incubated on a prepared sub-percolated gold/graphene/$SiO_2$/Si FETs. Oligonucleotides that were anti-sense and non-sense to the thiolated ssDNA probes were acquired and exposed to the ssDNA-S-gold/graphene FET to validate the biosensor scheme. Then, the response to synthetic SARS-CoV-2 mRNA was studied. The sequence space of ssDNA probes is vast, enabling surveillance of nearly any RNA transcript upon recombination of the thiolated ssDNA sequence. The results described hereafter suggest that combining ssDNA sequence space with cleanly fabricated graphene biosensors demands more scientific and, ultimately, commercial attention for augmenting modern transcriptomics.

## Results and Discussion

### Sheet Resistance and Physical Characterization of Ultrathin Gold on Graphene FET

As observed in **1a**, increasing the thickness of ultrathin gold/graphene device was found to increase the sheet resistance up to $t_c$, whereafter the sheet resistance decreased markedly. These results build on the previous work of Tatarkin *et al.* by exploring films <5 nm.[30] The devices used for biosensing experiments had a thickness of gold just below the $t_c$ to mitigate any drop in sheet resistance and maintain sensitivity. Physical characterization was then implemented as shown in **Figure 1b**.

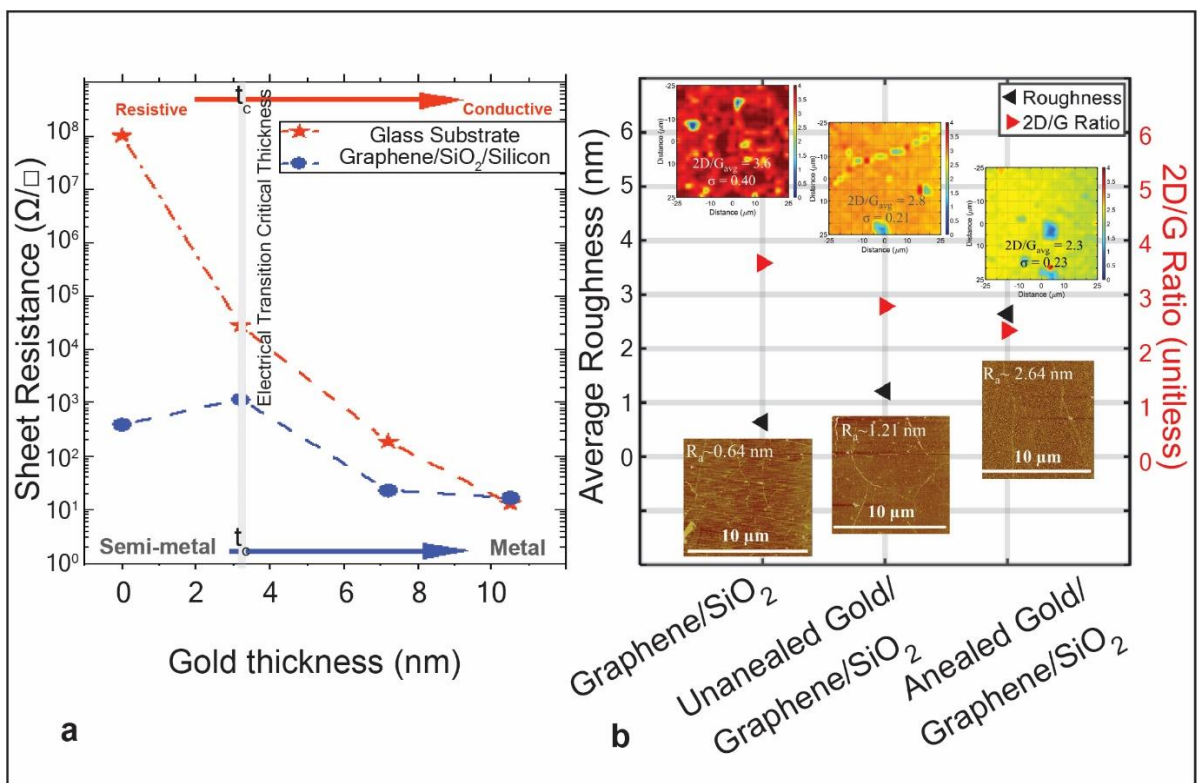

**Figure 1. (a)** thickness of Au from electron beam deposition versus the sheet resistance on glass slide (red stars **∗**) and on the graphene devices (blue circles •). Electrical transition thickness is given as $t_c$. **(b)** Raman spectral mapping and Atomic fore microscopy topography. The x-axis, left to right, is graphene/SiO$_2$/p-type Si, before and after annealing <3 nm thick Au/graphene/SiO$_2$/p-type Si.

Commercial monolayer graphene/SiO$_2$/p-type Si was interrogated with atomic force microscopy (AFM) Raman spectral mapping (**1b**). Then, immediately after electron beam deposition of <3 nm Au onto the graphene material, the characterization was repeated and the sample subsequently annealed, and characterized again. The top three insets are the Raman spectral mapping results, whereas the bottom three insets are the atomic force microscopy (AFM) topography. The left and right y-axis plot the average roughness (nm) and Graphene phonon intensity ratios (2D/G), respectively.

AFM and Raman spectral mapping were used to describe the topology and quality of the graphene devices. The starting sample, monolayer graphene/SiO$_2$/p-type Si had an exceptional average 2D/G intensity ratio of 3.6 (σ = 0.40) and an average roughness $R_a$ of 0.64 nm. Deposition of 2.5 nm of gold resulted in a 2D/G ratio of 2.8 (σ = 0.21) and an $R_a$ of 1.21 nm. Then, an annealing step was implemented to facilitate recrystallization of the gold on the graphene. This yielded an even lower 2D/G ratio of 2.3 (σ = 0.23) with a higher $R_a$ of 2.64 nm.

### Fabrication and Characterization of Graphene FET

The graphene biosensor is derived from a graphene FET fabricated on commercial four-inch graphene/90 nm SiO$_2$/p-type Si (see **Materials and Methods** for FET fabrication) and the functionalization steps are outlined from **Figure 2a** to **2b**. Briefly, the graphene FET has an ultrathin layer of gold deposited from electron beam deposition and is then incubated with reduced, thiolated ssDNA. The sequence and structure of the 20-mer ssDNA probe is given in **2c** (*5'-GGCCAATGTTTGTAATCAGT-3'-thiol*). Alafeef *et al.* have used this optimized sequence in their graphene nanoplatelet scheme and validated it for sensing SARS-CoV-2 N gene mRNA.[27] A cross-sectional view of the graphene biosensor is displayed in **2d**. Here, an acrylic well is bonded and used to facilitate incubation of reduced, thiolated ssDNA and liquid gating during characterization of and experimentation with the biosensor.

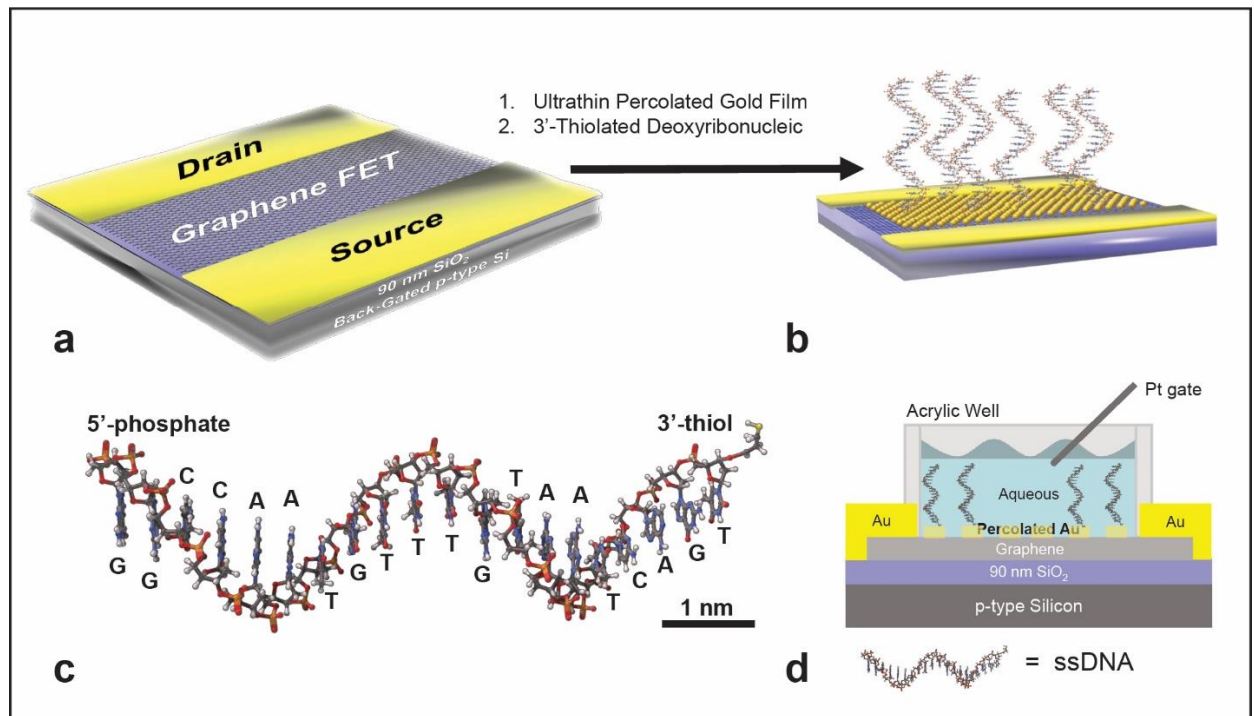

**Figure 2**. **(a)** metalized graphene/90 nm $SiO_2$/p-type Si. Two steps to produce **(b)**: (1) electron beam deposition of ultrathin percolated gold film and (2) incubation with reduced, thiolated ssDNA. **(c)** atomic model of the 20-mer ssDNA sequence used in this work. **(d)** cross-section of the graphene biosensor.

An exemplar gold/graphene FET was electrically characterized throughout the biochemical conjugation in **Figure 3a** and **3b** yielding two-terminal and back-gated three-terminal characteristics, respectively. The green lines show the initial electronic properties, whereas the blue lines show how the electronic properties change after incubating with reduced, thiolated ssDNA at 4 °C overnight. The initial Dirac voltage is +1.6 volts, indicating a relatively low level of p-type doping. Reduction and incubation of the ssDNA on the gold/graphene FET are outlined further in the **materials and methods**.

After this incubation, the devices were washed thoroughly by repeated, gentle medium exchange with a pipette. The washing resulted in the purple lines in **Figure 3**. It was found that the final Dirac voltage was -6.4 volts, indicating the device had become slightly n-type. The final, washed state suggests that the functionalization of thiolated ssDNA donates electrons to the gold/graphene FET. It should be noted that the 3'-thiolate is nucleophilic.

This is interesting as the ssDNA is net negatively charged in solution and it would be anticipated the graphene material would generate more holes to maintain charge neutrality. There may be competing electron induction events likened to alkyl chains inductively donating electron density to benzene and other polyaromatic hydrocarbons. If the functionalization donates electron density as observed in **3b**, then the hybridization event between anti-parallel, complementary polynucleotides may compensate the initial donation by withdrawing some electron density from the ssDNA probe, causing the device to become more p-type. This is likely compounded further by the introduction and localization of more net negatively charged polynucleotide to the graphene surface.

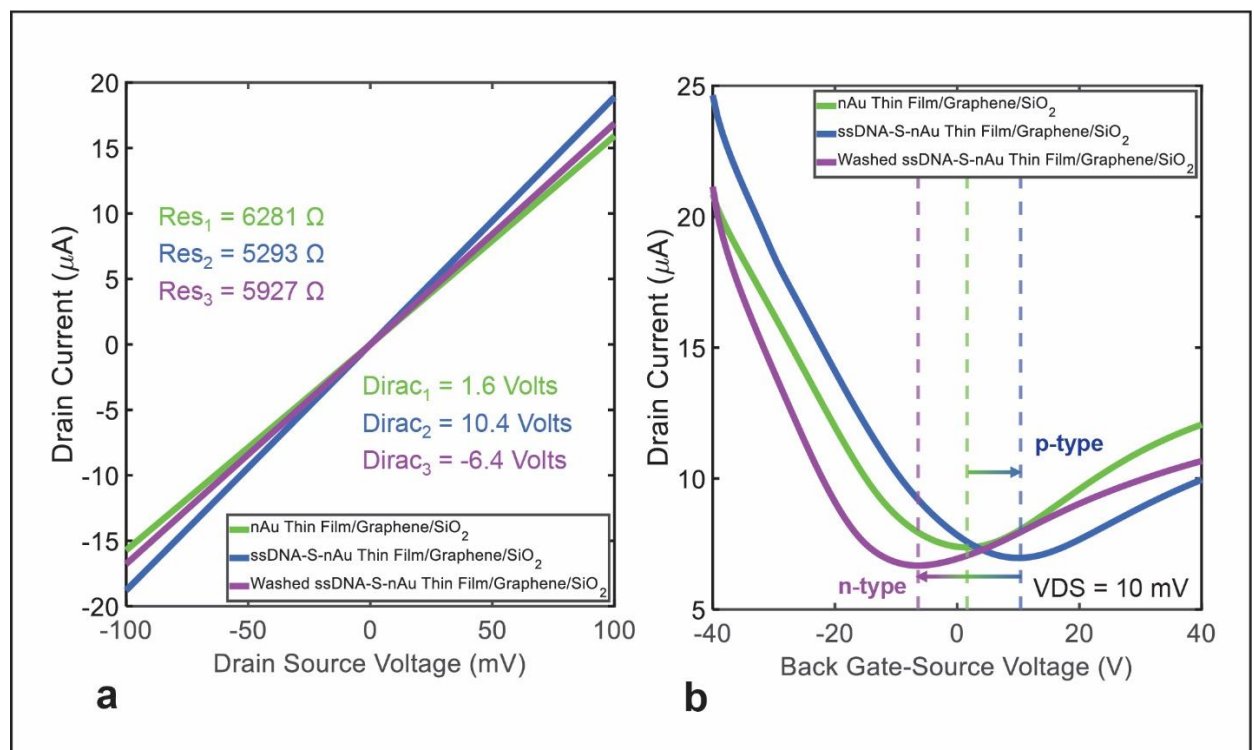

**Figure 3.** (**a**) two-terminal measurement of <3 nm of Au/graphene/SiO$_2$/p-type Si (green) throughout thiolated ssDNA functionalization (blue) and washing (purple). (**b**) Back-gated three-terminal measurement of <3 nm of Au/graphene/SiO$_2$/p-type Si (green) throughout ssDNA functionalization (blue) and washing (purple). The resistances and Dirac voltages for each step are inset in (**a**). The transfer characteristics were obtained with a drain-source voltage of 10 mV.

### Optically Characterizing Gold-Thiol Chemistry

To validate the gold-thiol chemistry, ultraviolet/visible (UV/Vis) spectroscopy of 5 nm diameter gold spheres before and after reacting with reduced, thiolated single stranded DNA (ssDNA) in molecular biology grade deionized water. The UV/Vis spectra were also investigated when a complementary, anti-sense ssDNA was introduced to the ssDNA-gold nanoparticle conjugate. UV/vis experiments were done to conclude on the efficacy of reacting 3'-thiolated ssDNA with gold nanomaterials, and the effect of introducing a hybridization partner (*i.e.*, the anti-sense strand of said 3'-thiolated ssDNA) on the nanogold surface plasmon resonance (SPR) frequencies.

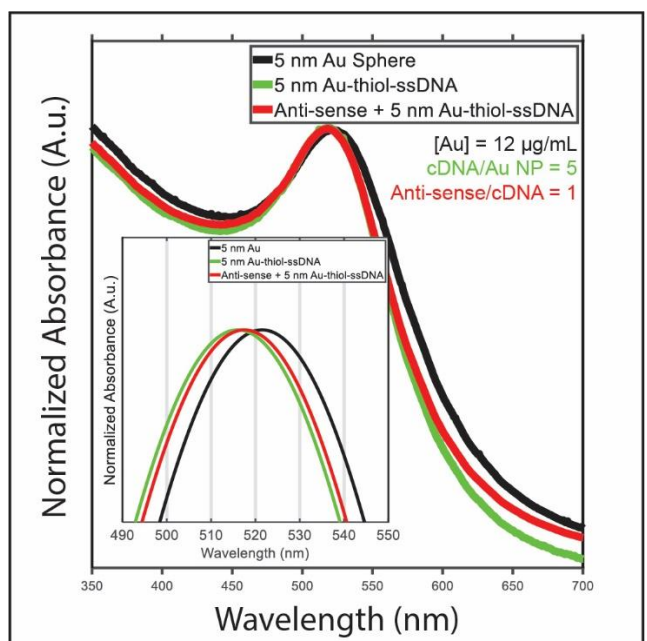

**Figure 4**. Ultraviolet/visible spectroscopy of 5 nm diameter Au spheres (black), ssDNA-conjugated 5 nm Au spheres (green), and the conjugated ssDNA-Au following incubation with the anti-sense sequence (red).

Specifically, the black line represents 5 nm diameter gold spheres suspended in ultra-pure water with no conjugation (stabilized by citrate, however). The peak resonance here is 521 nm and deviates from reported values for the SPR of 5 nm gold spheres likely due to aggregation, however, does not detract from the conclusions made because all measurements used the same stock of sonicated gold nanoparticle suspension. The green line represents what happens when reduced, 3'-thiolated ssDNA are conjugated to the nanogold spheres. Upon conjugation, the nanogold sphere peak resonance blue shifts to 516 nm ($\Delta\lambda_{ssDNA}$ = 5 nm). When introducing the anti-sense strand to the conjugated probes, the red line is measured. Here, a red shift to 518 nm is observed. These UV/vis spectroscopy results show that the nanogold spheres peak SPRs change when incubated with reduced, thiolated ssDNA and, subsequently, when the ssDNA-nanoparticle conjugate is introduced to an anti-sense polynucleotide.

The blue shift observed here contrasts to previous literature where pegylated gold nanoparticles are conjugated with thiolated ssDNA and experience a redshift.[31] This reported redshift could be due to a more pronounced increase in size where the poly(ethylene glycol) stabilizes the ssDNA away from the gold into the solvent. The gold nanoparticles used here were citrate-stabilized and experienced a blue shift upon conjugation possibly due to electron donation/electrostatic adsorption that was more dominant than an increase in effective diameter. Characterization of the dielectric constant and degree of aggregation of the conjugated-gold nanoparticles would offer more insight into these frequency disparities.

## Liquid-Transfer Characteristics of Graphene Biosensor: Anti-sense ssDNA

**Figure 5** investigates the graphene biosensor's liquid-gated three terminal response to anti-sense 20-mer ssDNA. This was done by using an acrylic well for initializing the device with 20 μL of fresh, filtered DI water and submerging a platinum electrode. After initialization, the devices were liquid-gated every time a droplet was added to the well ($V_{Droplet} = 2$ μL). The sequence of droplets started with 0 copies of ssDNA (blank water), and after assuring the transfer characteristics were stable, the droplets were doped with diluted anti-sense ssDNA. The domain of anti-sense ssDNA concentration in bulk liquid was [100 aM, 1 nM] and the droplets were added from lowest to highest concentration in decade intervals (see **Materials and Methods** for preparing dilutions).

**Figure 5**. (**a**) liquid-transfer characteristics of a negative-probe control (Au/graphene/SiO$_2$/p-type Si without probes) after exposure to serially concentrated dilutions of anti-sense ssDNA. (**b**) liquid-transfer characteristic of a fully

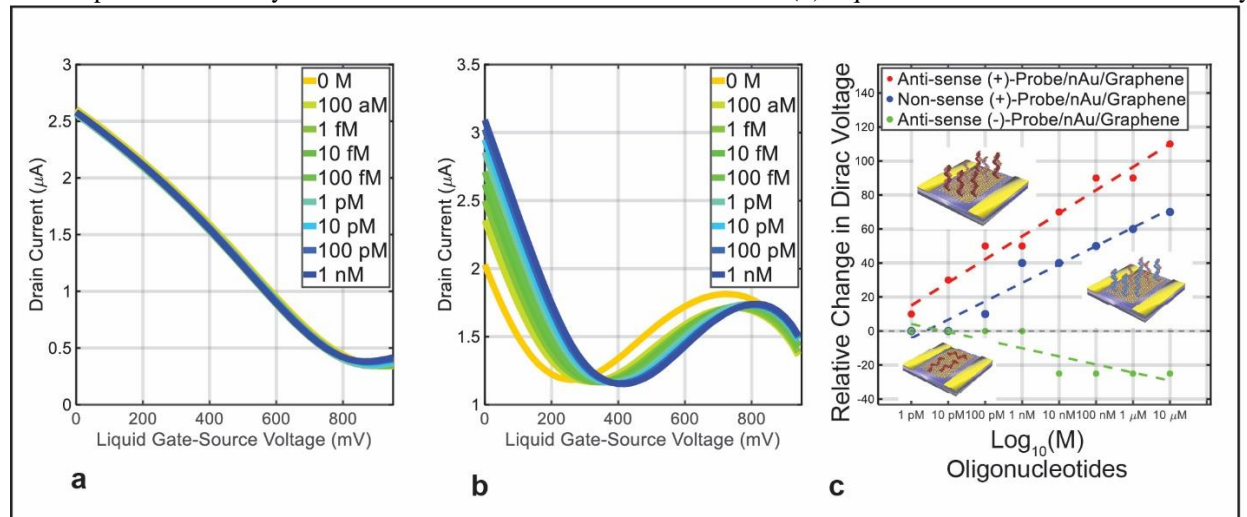

functionalized ssDNA-S-Au/graphene/SiO$_2$/p-type Si after exposure to serially concentrated dilutions of anti-sense ssDNA. (**c**) relative change in the Dirac Voltage versus concentration of serially concentrated dilutions of anti-sense ssDNA and non-sense ssDNA (not shown here). Insets represent system being measured: red for anti-sense oligonucleotide, blue for non-sense oligonucleotide, and the green data represents a negative-probe control.

**Figure 5a** shows the transfer characteristics of a negative-probe control (Au/graphene/SiO$_2$/p-type Si without ssDNA probes), whereas **5b** shows the liquid-transfer characteristic of a incubated, washed ssDNA-S-Au/graphene/SiO$_2$/p-type Si after exposure to serially concentrated dilutions of anti-sense ssDNA. **5c** plots the relative change of the Dirac Voltage versus the logarithm of the molarity of oligonucleotide where the initial Dirac voltage is extracted from the initialization with water blanks. The green data in **5c** shows the negative-probe control, whereas the blue data is the data for a non-sense ssDNA exposure experiment (see **Supplementary Information**).

The negative-probe control exhibited a different trend than the functionalized graphene biosensors, suggesting the ssDNA probes efficaciously functionalize the gold/graphene FET. The non-sense oligonucleotide is not expected to hybridize completely with the ssDNA-S-Au-graphene device, but because of the short length of 20-mers and the moderate guanidine-cytosine content it is possible there is molecular association between the non-sense ssDNA and the biosensor. The anti-sense ssDNA had the highest response, suggesting a stronger association than the non-sense. Overall, the Dirac voltages shifted to higher liquid-gate-source voltages, indicating an increase in p-type doping when exposed to anti-sense ssDNA.

### Liquid-Transfer Characteristics of Graphene Biosensor: Synthetic SARS-CoV-2 RNA

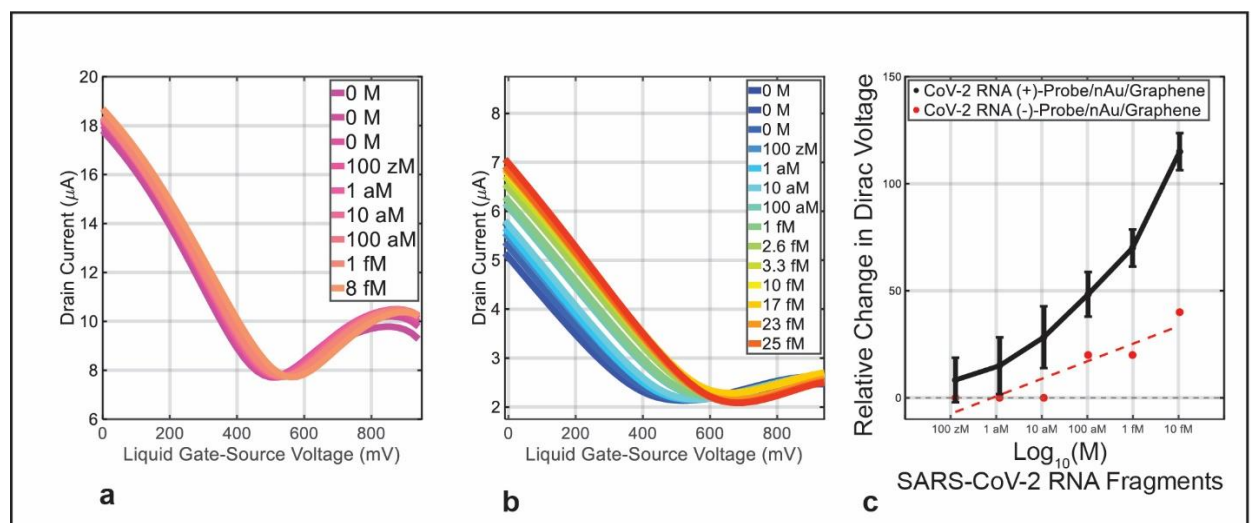

**Figure 6**. (**a**) liquid-transfer characteristics of a negative-probe control (Au/graphene/SiO$_2$/p-type Si without probes) after exposure to serially concentrated dilutions of synthetic SARS-CoV-2 RNA. (**b**) liquid-transfer characteristic of a fully functionalized ssDNA-S-Au/graphene/SiO$_2$/p-type Si after exposure to serially concentrated dilutions of synthetic SARS-CoV-2 RNA. (**c**) relative change in the Dirac Voltage versus concentration of serially concentrated dilutions of synthetic SARS-CoV-2 RNA. The functionalized biosensors were done in triplicate.

After the results were generated for **Figure 5**, the graphene biosensors were then tested for efficacy against their target: SARS-CoV-2 N gene mRNA. Synthetic SARS-CoV-2 genome (MN908947.3) was commercially acquired. The same protocol was followed as for the anti-sense ssDNA, however, the concentration domain for the SARS-CoV-2 RNA was [100 zM, 25 fM] resulting in **Figure 6**. The molar mass of the RNA genome is significantly larger than the 20-mers used in **Figure 5**. The negative-probe control (**6a**) had Dirac voltages shifting slightly positive after exposure. This is a different trend than the negative-probe results for the anti-sense ssDNA in **5a** and may be explained by the differences between the size of the ssDNA 20-mer and the kilomer-scale SARS-CoV-2 genome.[33]

The functionalized graphene biosensors, however, showed a larger, repeatable response (**6b**) than the negative-probe control. The scale of the shift in Dirac voltage of the functionalized graphene biosensor for anti-sense ssDNA and synthetic SARS-CoV-2 RNA is similar, suggesting that hybridization events are occurring on the surface of the ssDNA-S-gold/graphene FET. These hybridization events localize the charged polynucleotide molecules enabling electrostatic coupling with the gold/graphene FET surface.

## Conclusions

Coating graphene FETs with ultrathin sub-percolated gold films yields a platform facilitating gold-thiol conjugation of biomolecules. The sheet resistance of the graphene material was characterized with various gold thicknesses deposited and revealed an electrical transition critical thickness ($t_c$). The sheet resistance increased with gold thickness up to $t_c$ and declined significantly with thicknesses greater than $t_c$. The thickness of gold used in the graphene biosensors was <3 nm and was physically characterized with Raman spectral mapping and AFM in **1b**.

The gold deposited on the graphene FETs decreased the 2D/G phonon intensities (ratios are positively correlated to in-plane vibration and graphene quality) and increased the average roughness ($R_a$). An annealing step further exaggerated the effect of the ultrathin gold film on the Raman and AFM signatures. Then, the thiolated ssDNA probes were shown to conjugate to gold nanoparticles by blue shifting the SPR frequency in **Figure 4**.

The blue shift observed here may be correlated to the final electronic state of the washed ssDNA-S-gold/graphene FET suggesting an n-type nature (**Figure 3b, green**). The plasmon resonance is related to the metals charge carrier density through the plasma frequency, the dielectric constant, and the size.[32] Hybridization of an anti-parallel complementary oligonucleotide, however, likely compensates the initial donation by inductively withdrawing electron density from the gold/graphene FET, resulting in a shift towards p-type doping when back-gated (**Figure 3b**). A red shift was observed in the UV/Vis of this scenario with gold nanoparticles (**Figure 4**).

Gating the ssDNA-S-gold/graphene FET through the liquid with platinum revealed the initial electronic state to be p-type. This difference from the back-gated transfer characteristics may be explained by interactions with water molecules and conjugated ssDNA at the gold/graphene surface as the liquid-transfer characteristic is measured but remains beyond the scope of this work. When the ssDNA-S-gold/graphene FETs are exposed to anti-parallel, complementary oligonucleotides (anti-sense ssDNA and synthetic SARS-CoV-2 RNA in **Figures 5** and **6**) and transfer characteristic's point of minimum conduction shifted to more positive liquid-gate voltages. Thus, the ssDNA-S-gold/graphene FET becomes more p-type as the molarity of complementary polynucleotides increases.

The anti-sense ssDNA sensitivity acts as a positive control system, as the hybridization is programmed by the sequence complementarity. **5b** shows the ssDNA-S-gold/graphene FET responds with p-type doping. This conclusion is bolstered by the observed transfer characteristics of the negative-probe control (gold/graphene FET) in **5a**. The non-sense ssDNA sensitivity is plotted in **5c** and shows some sensitivity, which is fair given the short nature of the oligonucleotides, and the moderate guanidine-cytosine content, are liable to off-target association.

The ssDNA-S-gold/graphene biosensor was then studied for sensitivity to synthetic SARS-CoV-2 RNA. This assay was completed in triplicate and shown to be comparatively different to the negative-probe control (**6a** and **6b**) and had a similar result to the ssDNA positive control system in **5b**, a system designed to hybridize. In this work, the limit of detection for SARS-CoV-2 mRNA is 1 aM (approximately 1 copies per μL in the bulk) in filtered, deionized water with a dynamic range of 4 orders of magnitude. The linear sensitivity, from 10 aM to 10 fM, of the Dirac voltage was approximately 22 mV/decade of SARS-CoV-2 RNA molarity. The performance of the system used in Alafeef *et al*., using gold nanoparticles and graphene nanoplatelets, produced merits like the ultrathin gold/graphene/SiO$_2$ FETs employed here.[27]

The ultrathin gold/graphene FET is a promising approach to building a transcriptomics platform for incredulously dilute RNA transcripts because the measurands relate to the transport properties of the device. The platform modularity comes from the sequence of the thiolated ssDNA and can be reliably conjugated by sulfur-gold covalency. If a gene has been sequenced, probes can be engineered to target it. Using multiple ssDNA probes for the same target could augment the selectivity of the sensor and one could imagine the biosensor activity being a function of the mole fraction of the ssDNA probes (Alafeef *et al*. does exactly this).[27] Additionally, the probe density could be engineered through mixture with other thiol molecules to block gold atom reaction sites.

The results of this report suggest the ssDNA-S-gold/graphene FET has great potential as an ultra-sensitive, programmable, and cost-efficient biosensor for gene transcripts. To further validate this work for commercial sensing application, the ssDNA-S-gold/graphene FET should be tested against clinical and/or gamma-irradiated samples containing the SARS-CoV-2 virion and compared to the ideal sensing conditions studied here. Other strains could also be studied, like MERS-CoV and SARS-CoV-1. If further validated, this biosensing scheme has the potential to augment contemporary transcriptomic methods, like microarrays, RNA-seq, and polymerase chain reaction technologies. Understanding the complex regulatory events occurring at the RNA level may solve some of the greatest modern medical challenges.

## Materials and Methods

### Reagents

Four-inch monolayer graphene/90 nm of $SiO_2$/p-type silicon was purchased commercially from Grolltex. Stainless-steel shadow masks were obtained from OSH Stencils. Reagents were purchased commercially from well-known vendors. The thiolated-ssDNA, the anti-sense ssDNA, and non-sense ssDNA were purchased from GenScript in a dry form and stored in the freezer (-20 $^{o}$C). All sensing studies were performed at room temperature (~25 $^{o}$C). All water used was HPLC-grade and filtered. Tris(2-carboxyethyl)phosphine hydrochloride (TCEP) was obtained from Millipore Sigma. Dry, synthetic SARS-CoV-2 RNA fragments (MN908947.3) were procured from Twist Bioscience (Part number 103925) and resuspended in water for experimentation. The dry, synthetic RNA was stored short-term in the freezer (-20 $^{o}$C) until ready for experiment. It is recommended that long-term storage in a freezer of freezer of -80 $^{o}$C. Fresh reagents were used for every experiment.

### Gold/Graphene FET fabrication

Four-inch graphene/90 nm of $SiO_2$/p-type silicon was acquired. Using a stainless-steel shadow mask, samples were metallized with 100 nm of Au with a sensing area of 2000 x 2000 μm$^2$ (L × W) via electron beam deposition with Angstrom deposition system. The sensing area was then covered with 2.5 nm of gold, followed by 9 nm of $Al_2O_3$ by electron beam deposition and shadow masking. Reactive ion etching with oxygen plasma was then employed to subtract uncovered graphene from the wafer surface. The wafer was then cleaved with a diamond tip pen for chip-scale processing. The $Al_2O_3$ was removed with basic etch using TMAH-based developer, rinsed gently with deionized water, and the resulting device subject to an annealing step under argon/hydrogen atmosphere. After annealing, an acrylic well was aligned by hand and bonded to the $SiO_2$ surface with curable adhesive that protected the source and drain contacts from contacting the solution volume. To access the p-type silicon for back-gating the transfer characteristics, a scratch to expose the Si was scored away from the sensing region with the diamond tip pen.

### 3'-Thiolasted ssDNA Functionalization of Gold/Graphene FET and the Performance

Thiolated ssDNA comes as a disulfide and requires reduction to free sulfhydryl prior to reacting with gold materials. From a dry powder, 0.5 nmol disulfide ssDNA was suspended in 10 μL of water (50 μM of disulfide form) and mixed with 10 μL of 3 mM aqueous TCEP. The new solution was mixed gently with the pipette and the resulting mixture allowed to incubate at room temperature (25 °C) for 30 minutes in the dark. At 30 minutes, 80 μL of water was added was added and mixed gently, resulting in 100 μL of reduced, thiolated ssDNA at 10 μM (assumes complete reduction yielding 2 molar equivalences of 3'-thiolated ssDNA from the disulfide form). From this mixture, 20 μL intervals were distributed to gold/graphene FETs bonded with acrylic wells. The devices were incubated with this volume at 4 °C overnight.

The well volumes did not evaporate completely, and the devices were washed thoroughly with fresh water. This was done by adding 20 uL of water, mixing gently with a pipette, and then removing 18 uL of the newly added volume. This washing step was repeated seven times by adding and removing 18 uL of water. Ideally, residual TCEP and unreacted ssDNA molecules were removed with this medium exchange series. The devices were then subject to serially diluted samples containing either anti-sense ssDNA or synthetic SARS-CoV-2 RNA. Dilutions were prepared by adding 10 μL of suspended polynucleotide to 90 μL of water. These dilutions were mixed thoroughly with the pipette prior to making the next serial dilution.

### Electrical Measurments

Electrical characteristics were attained with Keysight B2902A source meter and a probe station. All three-terminal measurements were done with a drain-source voltage of 10 mV. Back-gated and liquid-gated transfer characteristics were in the domain and polarity of [-40, +40] and [-1, +1], respectively. The measurements used 16 power line cycles. Liquid-gated measurements were made with clean platinum electrode.

### UV/Vis Spectroscopy

Ultraviolet/visible spectrum of gold nanoparticles (and their conjugates) were obtained with Molecular Devices SpectraMax Plus 384 microplate reader using 1 nm intervals on the domain of [250, 1000] nm. Gold nanospheres (diameter = 5 nm) were diluted to a stock of 12 μg/mL (approximately $10^{13}$ nanoparticles/mL). Using 198 μL of this dilution, 2 μL of 50 μM reduced (approximately $5 \times 10^{13}$ particles), thiolated ssDNA was added and mixed thoroughly by pipetting and allowed to incubate at room temperature for 30 minutes. After incubation, 2 uL of 50 μM of anti-sense ssDNA was mixed in by pipetting. The spectra were then collected for 5 nm diameter gold nanospheres, the ssDNA-S-gold nanosphere, and the anti-sense with the conjugated nanoparticle; measurements of these mixtures were made in parallel.

### Raman Spectroscopy

Horiba Raman spectroscopy microscope was used for generating Raman spectral maps from 625 separate spectra over 62.5 x 62.5 μm² ($L \times W$) area. Spatial intervals between measurement points was 2.5 μm in both principal directions describing the surface plane. These were the measurement parameters: excitation wavelength of 532 nm, exposure time of 9 seconds and 6 accumulations per point with a power of 3.2 mW. A grating of 1200 I/mm was used, and the spectrum centered at 1800 cm$^{-1}$.

## Acknowledgments

The authors are grateful and acknowledge the funding provided by Analog Devices, Inc., the processing, and characterization facilities made accessible by Boston University's Division of Materials Science and Engineering. Additionally, the graphene material supplier, Grolltex made building the biosensors described here possible.

## Supporting Information Available

Liquid transfer characteristics of non-sense ssDNA on ssDNA-S-gold/graphene/SiO$_2$ and for SARS-CoV-2 RNA replicates

## Author Information

**Corresponding Author**

**Nicholas E. Fuhr** – Division of Materials Science and Engineering, Boston University

**Authors**

**Mohamed Azize** – Division of Materials Science and Engineering, Boston University

**David J. Bishop** – Division of Materials Science and Engineering, Boston University

## Notes

The authors have no competing financial interest.